\documentclass[%
preprint,
 amsmath,amssymb,
 aps,
]{revtex4-2}

\usepackage{graphicx}
\usepackage{dcolumn}
\usepackage{bm}
\usepackage[noabbrev]{cleveref}
\usepackage[mathlines]{lineno}

\usepackage[dvipsnames]{xcolor}

\crefname{equation}{eq.}{eq.}
\crefname{Equation}{Eq.}{Eq.}

\newcommand{\br}{{\cal B}}

\newcommand{\piz}{\pi^0}

\newcommand{\EE}{e^+e^-}
\newcommand{\MM}{\mu^+\mu^-}

\newcommand{\ppbar}{p\bar{p}}

\newcommand{\psip}{\psi(2S)}
\newcommand{\psipp}{\psi(3770)}
\newcommand{\jpsi}{J/\psi}

\newcommand{\upsi}{\Upsilon(1S)}
\newcommand{\upsip}{\Upsilon(2S)}
\newcommand{\mev}{\mathrm{MeV}}

\newcommand{\gev}{\mathrm{GeV}}

\newcommand{\gevcc}{\mathrm{GeV}/c^2}
\newcommand{\invfb}{\mathrm{fb}^{-1}}
\newcommand{\invpb}{\mathrm{pb}^{-1}}

\newcommand{\fb}{\mathrm{fb}}

\newcommand{\nb}{\mathrm{nb}}

\begin{document}


\title{Impact of the interference between the resonance and continuum amplitudes on
vector quarkonia decay branching fraction measurements}

\author{Y.~P.~Guo}
\email{guoyp@fudan.edu.cn}
\affiliation{%
 Fudan University, Shanghai 200433, People's Republic of China
 }
\author{C.~Z.~Yuan}%
 \email{yuancz@ihep.ac.cn}
\affiliation{%
Institute of High Energy Physics, Beijing 100049, People's Republic of China 
}%

\date{\today}

\begin{abstract}

The measurement of the branching fraction of a heavy quarkonium decaying into light hadronic 
final state at $\EE$ colliders is revisited. In $\EE$ annihilation experiments, 
background contributions from the continuum amplitude and its interference with the resonance 
amplitude are irreducible.  These effects become more and more significant as the precision of 
experimental measurements improves. While the former can be easily subtracted with data taken 
off the resonance peak, the latter depends on the relative size and phase between the resonance 
and continuum amplitudes. Two ratios are defined to estimate the size of these effects, 
$r_{R}^{f}$ for the ratio of the contribution of the interference term to the resonance term and 
$r_{c}^{f}$ for that to the continuum term. 
We find that $r_{R}^{f}$ could be as large as a few percent for narrow resonances, and both 
$r_{R}^{f}$ and $r_{c}^{f}$ could be large for broad resonances. This 
indicates that the interference effect is crucial for the measurements
of the branching fractions
aiming at the percent level or better precision and needs to be measured or estimated
properly. 

\end{abstract}

\maketitle


\section{\label{sec:intro}Introduction}

Heavy quarkonium is a multiscale system that can be used to probe all regimes 
of quantum chromodynamics (QCD)~\cite{BrambillaN-EidelmanS-HeltsleyBK}; thus, it 
provides an ideal ground test for the understanding of QCD. The test can be done 
by comparing various observables predicted by theory and measured by experiments. 
One such observable is the hadronic decay width or equivalently, the hadronic decay 
branching fraction, which can be examined to find the patterns and properties of 
quarkonium decays. In principle,  the hadronic decay of heavy quarkonium can be 
calculated theoretically with some input from experimental measurements. 
However, rigorous calculations are still very limited; many of the studies of the quarkonium 
decays are based on phenomenological models and observations in experimental data.

An example of the quarkonia decay pattern is the long-standing puzzle in charmonium 
sector, namely, the ``$\rho-\pi$ puzzle''. 
In perturbative QCD (pQCD), the ratio $Q^{f}=\br[\psip\to f]/\br[\jpsi\to f]$ is predicted 
to be $Q^{f}=\br[\psip\to\EE]/\br[\jpsi\to\EE]\approx 12\%$~\cite{rho-pi-12-rule}, where 
$\br[\jpsi\to f]$ and $\br[\psip\to f]$ are the branching fractions of $\jpsi$ and $\psip$ decay 
into the same hadronic final state $f$.
Violation of this ``12\% rule'' was first observed by the Mark II experiment in 
$\rho\pi$ and $K^{*}\bar{K}$ decay modes~\cite{MarkII-rhopi-1983}. It was confirmed 
by other experiments in more vector-pseudoscalar (VP) decay modes as well as 
vector-tensor (VT) decay modes~\cite{rhopi-CERN2005}. The puzzle has not been 
solved although many theoretical explanations have been proposed~\cite{rhopi-review-2007}. 
A similar rule for the ratio in bottomium sector
$Q^{f}_{\Upsilon}=\br[\upsip\to f]/\br[\upsi\to f]\approx0.77$
is expected. The test of $Q^{f}_{\Upsilon}$ from measurements is
inconclusive~\cite{upsilon-decay-belle,upsilon-belle2,upsilon-cleo}. 

Experimentally, vector quarkonia can be produced from $\EE$ and $p\bar{p}$ 
annihilation processes or heavier hadron decays; the branching fraction can be 
measured using subsequent decays. Among them, $\EE$ annihilation experiments 
are the most important contributors as they provide a clean experimental environment, 
and the vector quarkonia can be copiously produced at rest. Most of the hadronic 
decay branching fractions of charmonium states were measured with data samples 
collected by the CLEOc, BESII, and BESIII experiments, and the typical precision is 
a few tens of percent~\cite{pdg2008}. The hadronic decays of $\upsi$ and $\upsip$ 
were measured by the Belle and CLEO experiments~\cite{pdg2020}.
The best precision is several percent. The BESIII and Belle II are currently running 
$\EE$ experiments. At the BESIII experiment, 10 billion $\jpsi$ events and 3 billion $\psip$ 
events have been accumulated,  and $20~\invfb$ $\psipp$ data are expected before 
2024~\cite{bes3-whitepaper}. The sizes of the data samples are at least 6 times larger 
than those used in previous BESIII measurements,  and dozens of times larger than 
those in the CLEOc and BESII measurements. At the Belle II experiment,  about $500~\invfb$ 
data for each vector bottomnium state are planned~\cite{belleII-physics}, which are 
tens (hundreds) of times larger than those from the Belle (CLEO) experiment.
Therefore, the precision of the branching fraction of vector quarkonia decay 
can be improved significantly.

The branching fractions of hadronic decays measured in $\EE$ colliders have an 
unavoidable background contribution, $i.e.$, the continuum process produced 
directly from $\EE$ annihilation, $\EE\to\gamma^{*}\to f$. 
The signal events observed in an experiment contain contributions from both 
resonance decays and continuum production, and, more importantly, the interference 
between them. The importance of the continuum amplitude and the interference effect 
has been pointed out in Ref.~\cite{WYMZ-inter-2004}, but 
in most of the measurements, this was not taken into account properly due to the absence
or low statistics of data sample in the off resonance region.
In principle, to determine the branching fraction of a resonance decaying into 
a specific final state, one needs to know the cross section of the continuum 
production as well as the relative phase between 
the resonance and continuum amplitudes. This can only be realized by measuring 
the cross sections at no less than three energies around the resonance peak, since we 
do not have reliable theoretical or experimental knowledge on the continuum cross section
or the relative phase for any hadronic final state. However, this was not done in previous 
measurements where vector quarkonia are 
produced directly from $\EE$ annihilation.
For most cases in which the continuum contribution was considered, 
it was estimated using data samples taken off the resonance and subtracted without 
considering the interference effect. Moreover, the possible bias 
from this treatment of the interference effect was not included in
the systematic uncertainties. In old-generation experiments, both the 
statistical and systematic uncertainties are more than $10\%$, the interference effect 
may be neglected since it is not dominant. When one performs high precision measurements, 
this effect could be larger than many other sources of systematic uncertainties; thus, it 
can not be neglected. 

In this paper, we first revisit the cross section formula for a hadronic final state produced 
in an $\EE$ collider at a resonance peak, then we quantitatively describe
the importance of the continuum contribution and its interference 
with the resonance contribution in the branching fraction measurement. 
Finally, additional experimental effects from radiative correction and beam energy 
spread have been addressed.

\section{Production of a hadronic final state in $\EE$ experiment}
\label{ch:formula}

A hadronic final state in $\EE$ colliders in the vicinity of a resonance $R$ is 
produced via the coherent sum of the resonance and continuum amplitudes. If 
we use $a_c^f(s)$ to denote the continuum amplitude for a certain exclusive final 
state $f$, and $a_{R}^{f}(s)$ for the resonance amplitude, the cross section can be 
written as 
\begin{align}\label{eq1:born-cs}
    \sigma_{\rm tot}^f(s)&=|a_{c}^f(s)+ e^{i\varphi}\cdot a_{R}^f(s)|^2 
    \equiv \sigma_{c}^{f}(s)+\sigma_{R}^{f}(s)+\sigma_{\rm int}^f(s),
\end{align}
where $\sqrt{s}$ is the center-of-mass (c.m.) energy and $\varphi$ is the 
relative phase between the two amplitudes. We use 
$\sigma_{c}^f(s)=|a_{c}^{f}(s)|^2$, $\sigma_{R}^f(s)=|a_{R}^{f}(s)|^2$, and $\sigma_{\rm int}^f(s)$ 
to denote the cross sections of the continuum process, the resonance process, 
and the interference term, respectively. 
The resonance amplitude $a_{R}^f(s)$ is parametrized as 
\begin{equation}\label{eq:aR}
    a_{R}^f(s)=\frac{\sqrt{12\pi\Gamma_{ee}\Gamma\br_f}}
    {s-M^{2}+i M \Gamma},
\end{equation}
where $M$ and $\Gamma$ are the mass and total width of the resonance, 
$\Gamma_{ee}$ is the partial width of $R\to \EE$, and $\br_f$ is the branching 
fraction of $R\to f$. 
Inserting~\Cref{eq:aR} into~\Cref{eq1:born-cs}, the cross section is expanded as 
\begin{align}\label{eq:born-cs-expand}
    \sigma_{\rm tot}^f(s) &= \sigma_{c}^{f}(s) 
    +\frac{12\pi \Gamma_{ee}\Gamma\br_f}{(s-M^2)^2+M^2\Gamma^2} + 2 \frac{\sqrt{\sigma_{c}^f(s)}\sqrt{12\pi\Gamma_{ee}\Gamma\br_f}}
    {(s-M^2)^2+M^2\Gamma^2}
   [(s-M^2)\cos\varphi+M\Gamma\sin\varphi].
\end{align}
If the data sample is taken at the energy of the resonance mass, {\it i.e.}, 
$s=M^2$,~\Cref{eq:born-cs-expand} can be simplified as 
\begin{equation}\label{eq:born-cs-simplify}
    \sigma_{\rm tot}^f(s) = \sigma_{c}^{f}(s) +\frac{12\pi \br_{ee}\br_f}{M^2} 
    + 2 \frac{\sqrt{\sigma_{c}^f(s)}\sqrt{12\pi\br_{ee}\br_f}}{M}\sin\varphi,
\end{equation}
where $\br_{ee}=\Gamma_{ee}/\Gamma$ is the branching fraction of $R\to \EE$.

\begin{figure}[htbp]
    \centering
    \includegraphics[width=0.45\textwidth]{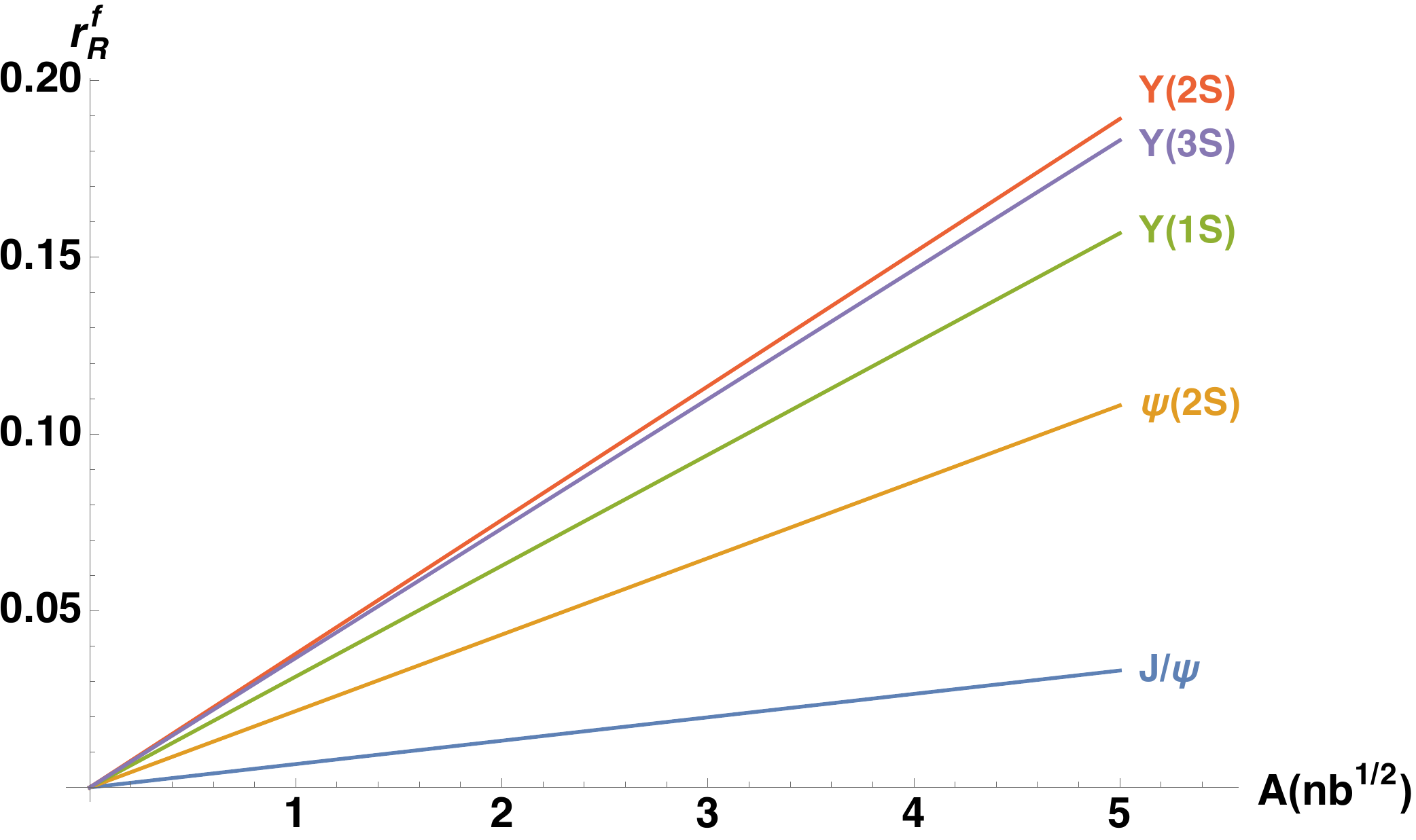}
    \includegraphics[width=0.45\textwidth]{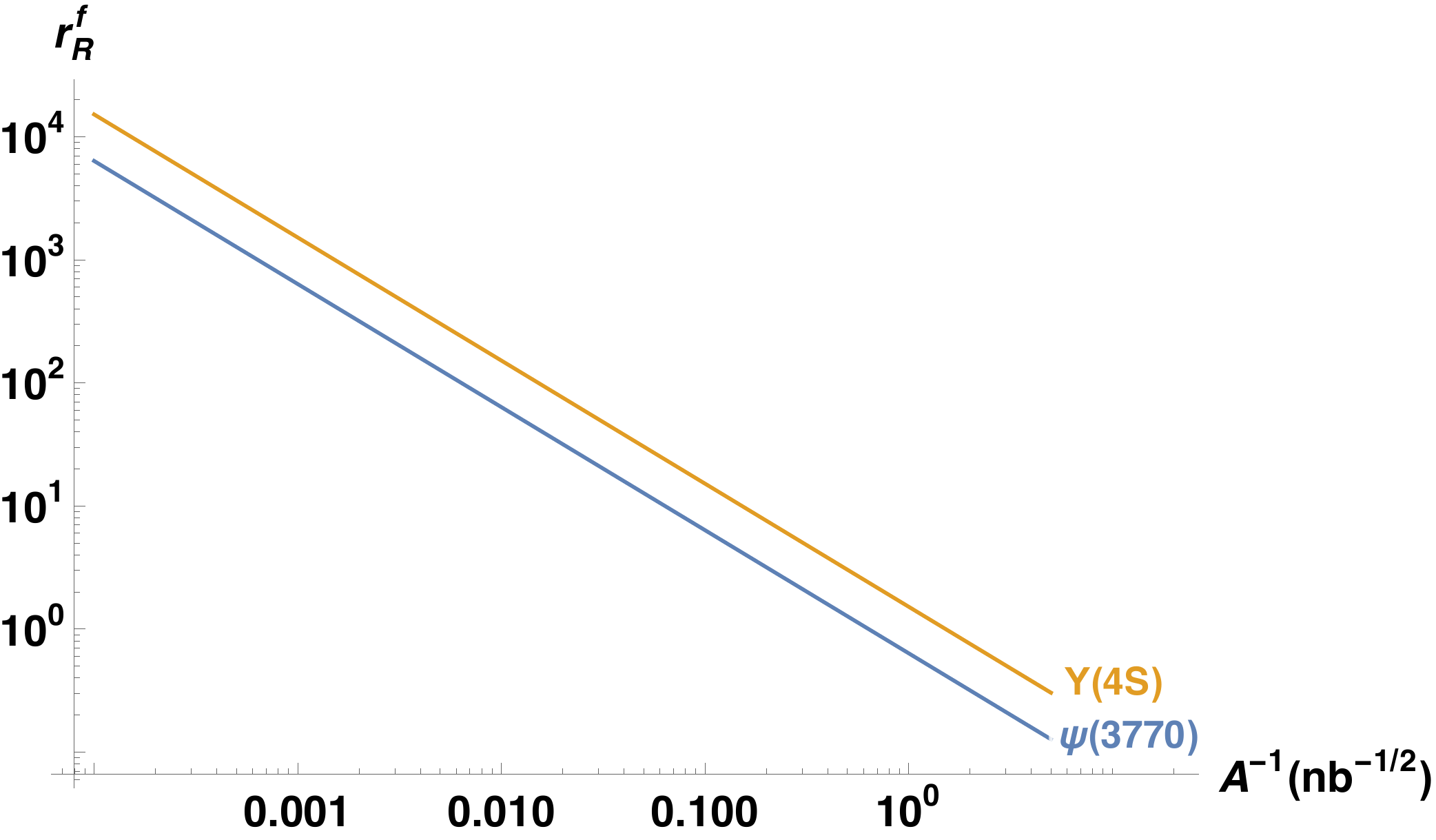}\\
    \includegraphics[width=0.45\textwidth]{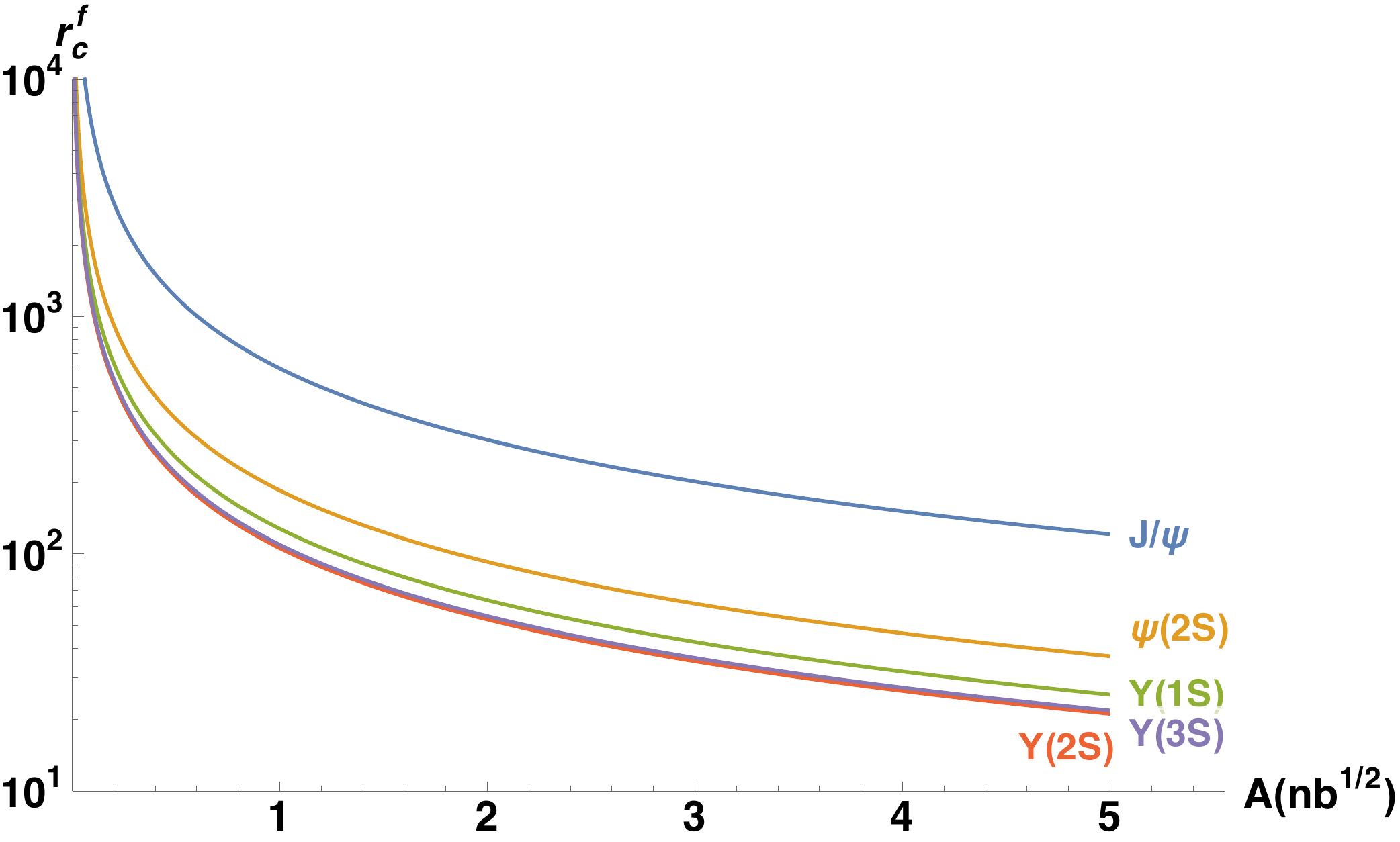}
    \includegraphics[width=0.45\textwidth]{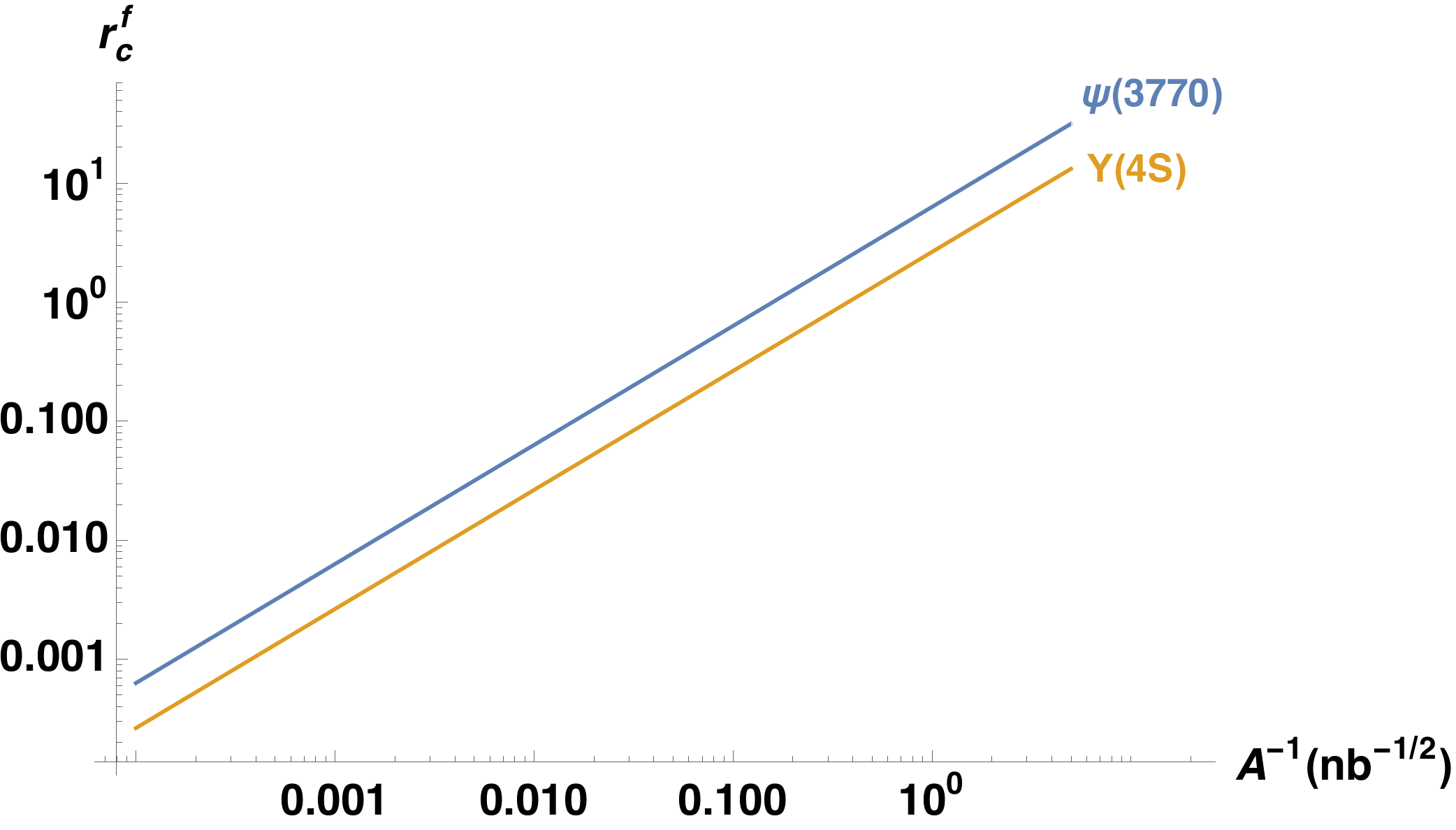}
    \caption{The dependence of $r_{R}^{f}{}^{\rm max}$ (top) and 
    $r_{c}^{f}{}^{\rm max}$ 
    (bottom) on $A$ for different resonances. The left column shows the 
    result for narrow resonances where the $x$-axis is set to $A$, and the 
    right column shows the result for broad resonances where the $x$-axis 
    is set to $A^{-1}$.}
    \label{fig:r-vs-A}
\end{figure}

Based on~\Cref{eq:born-cs-simplify}, we define two ratios, $r_{R}^{f}$ 
and $r_{c}^{f}$, representing the ratio of cross section from the interference 
term with respect to the resonance and continuum term, respectively,
\begin{align}\label{eq:r}
    r_{R}^{f} & \equiv \frac{\sigma_{\rm int}^{f}(s)}{\sigma_{R}^{f}(s)} = 
    \frac{2}{\hbar c}\sqrt{\frac{\sigma_{c}^f(s)}{\br_{f}}}
    \frac{M}{\sqrt{12\pi\br_{ee}}}\sin\varphi
    \equiv \frac{2}{\hbar c} A B\sin\varphi,\\ \nonumber
    r_{c}^{f} & \equiv \frac{\sigma_{\rm int}^f(s)}{\sigma_{c}^f(s)} = 
    2{\hbar c}\sqrt{\frac{\br_{f}}{\sigma_{c}^f(s)}} 
    \frac{\sqrt{12\pi\br_{ee}}}{M}\sin\varphi
    \equiv 2 {\hbar c} A^{-1} B^{-1} \sin\varphi.
\end{align}
Factor $A=\sqrt{\sigma_{c}^f(s)/\br_{f}}$ can be calculated 
once the cross section of the continuum process and the branching fraction 
are measured, and factor $B=M/\sqrt{12\pi\br_{ee}}$ is a constant
depending on the resonance parameters. Usually the cross section is measured 
in the unit of barn and $M$ in the unit of $~\gevcc$, so the conversion 
constant $\hbar c$ is added in the denominator (numerator) for 
$r_{R}^{f}$ ($r_{c}^{f}$). From~\Cref{eq:r}, it is obvious that the magnitudes
of the ratios reach maxima when the relative phase $\varphi$ is $\pm 90^{\circ}$. 
The dependence of $r_{R}^{f}{}^{\rm max}$ and $r_{c}^{f}{}^{\rm max}$
on $A$ is illustrated in Fig.~\ref{fig:r-vs-A} for narrow resonances and broad
resonances separately. 

\section{Narrow resonances}\label{ch:narrow-res}

Resonances with masses below open heavy flavor threshold, such as 
$\jpsi$, $\psip$, $\upsi$, $\upsip$, and $\Upsilon(3S)$, are narrow, and 
the production cross section in $\EE$ collision is much larger than the 
continuum cross section. Using $M$, $\Gamma$, and $\Gamma_{ee}$ values from 
PDG~\cite{pdg2020}, the total cross sections for the five resonances at peak 
positions can be calculated with 
\begin{equation}\label{eq:cs-res}
    \sigma_{R}(s)=\frac{12\pi\Gamma_{ee}\Gamma}{(s-M)^2+M^2\Gamma^2}=\frac{12\pi\br_{ee}}{M^2}.
\end{equation}
The inclusive hadronic cross sections of the continuum process can be estimated
by using the $\mathcal{R}$ value 
$\mathcal{R}(s)=\sigma_c(\EE\to hadrons)/\sigma(\EE\to\MM)$ and 
$\sigma(\EE\to\MM)=(4\pi\alpha^2)/(3s)$. With the $\mathcal{R}$ values from 
the latest BESIII measurement in the vicinity of $\jpsi$ and 
$\psip$~\cite{BESIII-R}, and from PDG for $\upsi$, $\upsip$, and 
$\Upsilon(3S)$~\cite{pdg2020}, the inclusive hadronic cross sections of the continuum process
at the five resonances peak positions are calculated and listed in 
Table~\ref{tab:born-cs-narrow-res}, together with the cross sections of
the resonance process and the $B$ factor. 
The absolute value of $\sigma_{c}$ is 3 or 4 orders of magnitude 
smaller than $\sigma_R$, while $\sigma_{\rm int}$ can be sizeable depending 
on the relative phase $\varphi$, as shown in the following example. 
So for narrow resonances, $r_{R}^{f}$ will be used to characterize 
the size of the interference effect. 

\begin{table*}[htbp]
\caption{The cross sections of the resonances ($\sigma_{R}$), the 
continuum process ($\sigma_{c}$), and the factor $B$
at the resonance peak positions.}
\label{tab:born-cs-narrow-res}
\begin{center}
\begin{tabular}{cccccc|cc}
\hline\hline
$\sqrt{s}~(\gev)$ & $M_{\jpsi}$  &  $M_{\psip}$  &  $M_{\upsi}$  &  $M_{\upsip}$ 
&  $M_{\Upsilon(3S)}$  &~$M_{\psipp}$~ & ~$M_{\Upsilon(4S)}$~\\\hline
$\sigma_{R}~(\nb)$ &~~~$91,404$~~~ &~~~$8,562$~~~&~~~$4,069$~~~&~~~$2,796$~~~
&~~~$2,984$~~~&~~~$9.9$~~~ &~~~$1.7$~~~\\ 
$\sigma_{c}~(\nb)$ &  $20.6$      &  $15.4$   &  $3.4$   &  $3.1$  & $2.9$ 
& $19.0$ & $2.8$ \\ 
$B~(\gevcc)$           &  $2.06$     & $6.74$   & $9.78$  & $11.8$  & $11.4$
& $198$ & $473$  \\
\hline\hline
\end{tabular}
\end{center}
\end{table*}

We use the branching fraction measurement of $\jpsi$ and $\psip\to\Lambda\bar{\Lambda}$ 
as an example to estimate the size of the interference effect.
The branching fractions of $\jpsi\to\Lambda\bar{\Lambda}$ 
and $\psip\to\Lambda\bar{\Lambda}$ were reported by the BESIII experiment 
using $1.3\times 10^{9}$ $\jpsi$ and $4.5\times10^{8}$ $\psip$ events, 
respectively~\cite{bes3-br-lambda}. A data sample at $\sqrt{s}=3.08~\gev$ 
($\sqrt{s}=3.65~\gev$) was collected with an integrated luminosity of $30~\invpb$ 
($44~\invpb$) to study the contribution from continuum process. No events passed 
the event selection from the sample collected at 
$\sqrt{s}=3.08~\gev$, and only six events survived from the sample collected 
at $\sqrt{s}=3.65~\gev$ which accounted for $0.34\%$ of the signal events 
selected from the $\psip$ sample. So in both cases, the continuum contribution was
neglected, and the branching fractions were determined. The measured branching 
fractions are summarized in Table~\ref{tab:ll-int}.  Lately, the BESIII experiment measured 
the cross section of $\EE\to\Lambda\bar{\Lambda}$ using data samples collected 
at c.m. energies 
from $3.51$ to $4.60~\gev$~\cite{bes3-cs-lambda}. It is found that the cross section 
can be described with a power-law function, $C\cdot(M_{\psipp}^2/s)^n$,
and the two parameters are determined to be $C=379\pm22$ and $n=8.8\pm0.4$.
Using the central values of these parameters, the continuum cross sections 
($\sigma_c^{\Lambda\bar{\Lambda}}$) at $\jpsi$ 
and $\psip$ peak positions are obtained and listed in Table~\ref{tab:ll-int}. 
Inserting all the numbers into~\Cref{eq:r}, the maximum values of $r_{R}^{f}$
can be read from Fig.~\ref{fig:r-vs-A} and are 
listed in the last column of Table~\ref{tab:ll-int}, which are $1.7\%$ and $2.6\%$ 
for $\jpsi$ and $\psip$, respectively; these can be compared with the 
quoted total systematic uncertainties of $1.7\%$ and $2.8\%$.

\begin{table*}[htbp]
\caption{The measured branching fractions ($\br^f$), the estimated cross sections 
from continuum contribution ($\sigma_{c}^f$), and the maximum 
value of $r_{R}^{f}$ at $\jpsi$ and $\psip$ peak position. }
\label{tab:ll-int}
\begin{center}
\begin{tabular}{ccccc}
\hline\hline
$~R~$ &~~~$\br[R\to\Lambda\bar{\Lambda}]~(10^{-4})$~~~ 
&~~~~~~$\sigma_{c}^{\Lambda\bar{\Lambda}}~(\nb)$~~~~~~&~$r_{R}^{f}{}^{\rm max}~(\%)$~\\\hline
${\jpsi}$  & $19.43\pm0.03\pm0.33$  &  $1.22\times10^{-2}$   & $1.7$ \\
${\psip}$  &~$3.97\pm0.02\pm0.12$   &  $0.57\times10^{-3}$  & $2.6$ \\
\hline\hline
\end{tabular}
\end{center}
\end{table*}

We find that the interference effect is surprisingly large compared with 
the precision that current experiments can reach for the two decays mentioned above. 
Since the hadronic decays of $\jpsi$ and $\psip$ have similar branching fractions
and the continuum cross sections of many final states are at a few to a few tens of
picobarn level, we expect a similar size of interference effect in other decay modes. 
In Fig.~\ref{fig:r-vs-A} we show $r_{R}^{f}{}^{\rm max}$ for $A=0\to 5$: this should cover most 
of the decay modes of these narrow resonances. Our result indicates that the interference
may change the measured branching fractions by subpercent to more than $10\%$ depending 
on different final states and different resonances, and the effect is more prominent for bottomonium states.
So this effect must be considered in evaluating the systematic uncertainties if it cannot be fully taken 
into account in the measurement of the decay branching fractions of narrow charmonium and bottmonium 
states. 

\section{Broad resonances}\label{ch:wide-res}

For resonances located above open heavy flavor threshold, such as the $\psipp$ 
and $\Upsilon(4S)$, the total cross sections of the resonance production
and continuum process are of the same level of magnitude, as shown in
Table~\ref{tab:born-cs-narrow-res}. However, the dominant decay modes of
these resonances are open heavy flavor final states; 
the decay into light hadronic final states is suppressed according to the OZI 
rule~\cite{ozi} and is only a tiny fraction of the total decay,
$(7^{+9}_{-8})\%$ for the $\psipp$ and $<4\%$ for the $\Upsilon(4S)$~\cite{pdg2020}. 
According to the available searches from previous 
experiments~\cite{BESIII:2021ftf,BESIII:ppbar,BESIII:pppi0,BESIII:2013ujm,BES:2007zan,CLEO:2005tkm,CLEO:2005zrs,Belle:2013hkg,Belle:2009rfa}, 
we have good reason to believe that the total decay rate to light hadrons should be 
at $1\%$ level for both $\psipp$ and $\Upsilon(4S)$. So the ratio
of the cross sections of the resonance decay and continuum production is at $0.5\%$ level.
It is worth noticing that the estimation of the total decay rate
to light hadrons for $\psipp$ ($1\%$) agrees with the theoretical prediction given in 
Ref.~\cite{He:2008xb}.
This is the reason we use $r_{c}^{f}$ to characterize the size of the 
interference effect in the case of broad resonances. In this case, the interference term 
introduces a deviation of the cross section measured at the resonance peak
from the smooth (almost constant) continuum cross sections in the vicinity 
of the resonance. The scale of the deviation depends on the factor $A$ 
defined in~\Cref{eq:r} for a certain resonance and the relative phase between 
the two amplitudes. The maximum deviation $r_{c}^{f}{}^{\rm max}$ is displayed
in Fig.~\ref{fig:r-vs-A}. 

The minor deviations, either higher or lower than the continuum cross sections, observed in
Refs.~\cite{BESIII:2021ftf,BESIII:2013ujm,BES:2007zan,CLEO:2005tkm,CLEO:2005zrs}
may have indicated nonzero $\psipp$ decays into light hadron final states,
but more data are needed to confirm the observations. 
In any case, the way of calculating the (upper limits of the) branching fractions 
without considering the interference as in the previous measurements
is incorrect, especially when the cross section at the resonance peak
is lower than that off the resonance, 
which reveals a destructive interference between the continuum and 
resonance amplitudes.
The measurements of $\EE\to \ppbar$~\cite{BESIII:ppbar} and 
$\ppbar\piz$~\cite{BESIII:pppi0} in the vicinity of the $\psipp$ peak 
show this effect and evidence for $\psipp$ decays to these final states. 

There are also cases where the branching fractions of resonance decay 
are large, and the cross sections from the continuum process, the 
resonance decay and the interference term are comparable. In 
these cases, all three components
are essential, and special precautions should be taken to obtain the correct
branching fraction of the resonance decay.

Again, we take $\psipp\to\Lambda\bar{\Lambda}$ as an example.
The total cross section of $\EE\to\Lambda\bar{\Lambda}$ at the $\psipp$ peak 
is measured to be $(562\pm 42)~\fb$~\cite{bes3-cs-lambda}. 
The cross section of the continuum process at $\psipp$ peak position is estimated as 
$379~\fb$ using the same formula as used in Sec.~\ref{ch:narrow-res}. 
Inserting the numbers into~\Cref{eq:born-cs-simplify}, the nominal 
value of the branching fraction of $\psipp\to\Lambda\bar{\Lambda}$ is calculated
to be in the range of $[1.79\times10^{-6},~1.88\times10^{-4}]$, depending on the value 
of $\varphi$. If the interference term is simply neglected, the branching fraction
is $1.8\times10^{-5}$, and could be very different from the true value depending on
the unknown $\varphi$. 
In Ref.~\cite{bes3-cs-lambda}, the cross section line shape is fitted with the 
coherent sum of $\psipp$ resonance and a power-law continuum term, and the relative 
phase is determined. In this case, the branching fraction can be determined exactly. 
The central value is $\br_{\rm con.}=2.4\times10^{-5}$ or 
$\br_{\rm des.}=1.44\times10^{-4}$~\cite{bes3-cs-lambda}. 
These are well covered by our estimated range above.

\begin{figure}[htbp]
    \centering
    \includegraphics[width=0.4\textwidth]{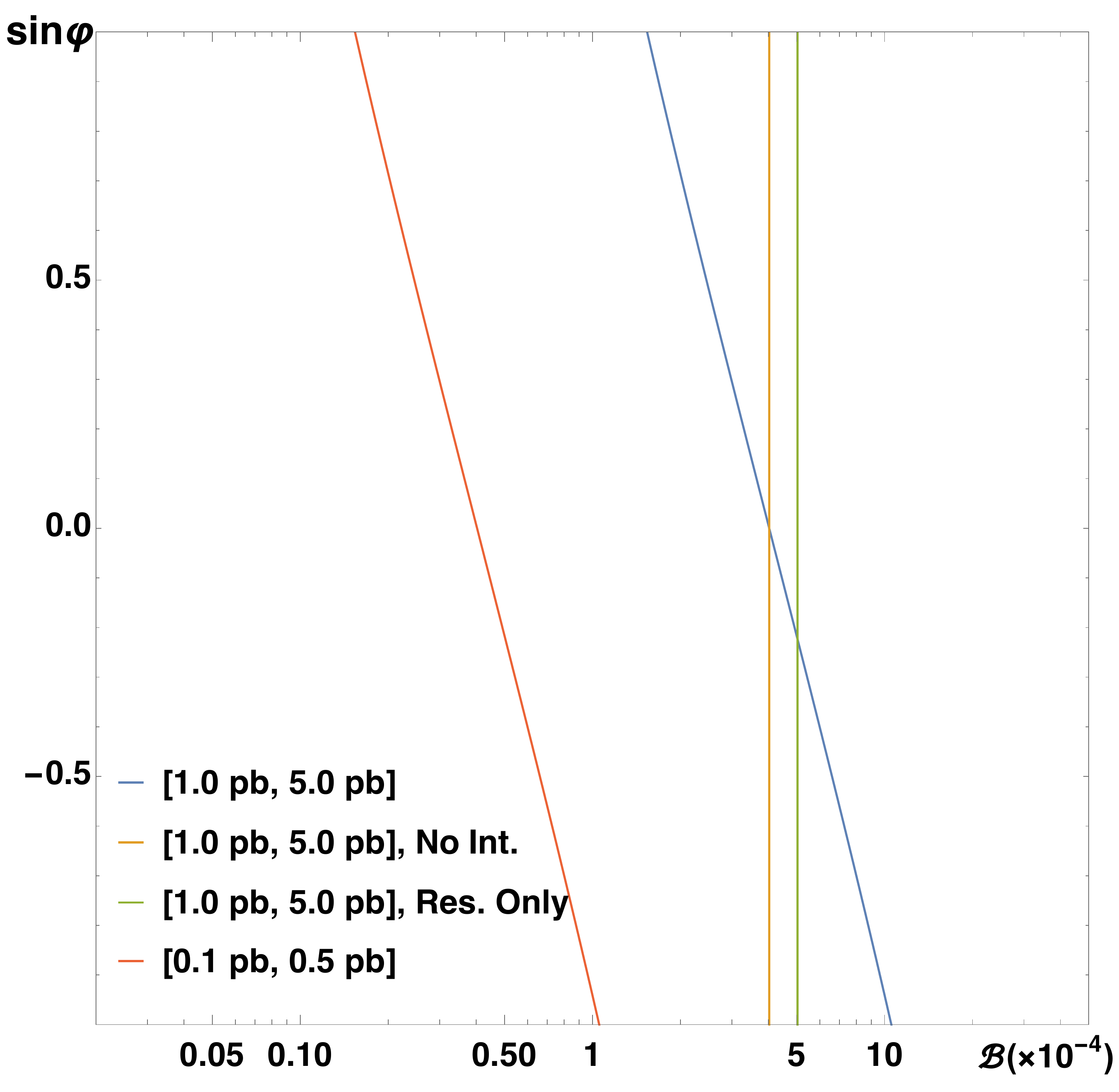}
    \includegraphics[width=0.41\textwidth]{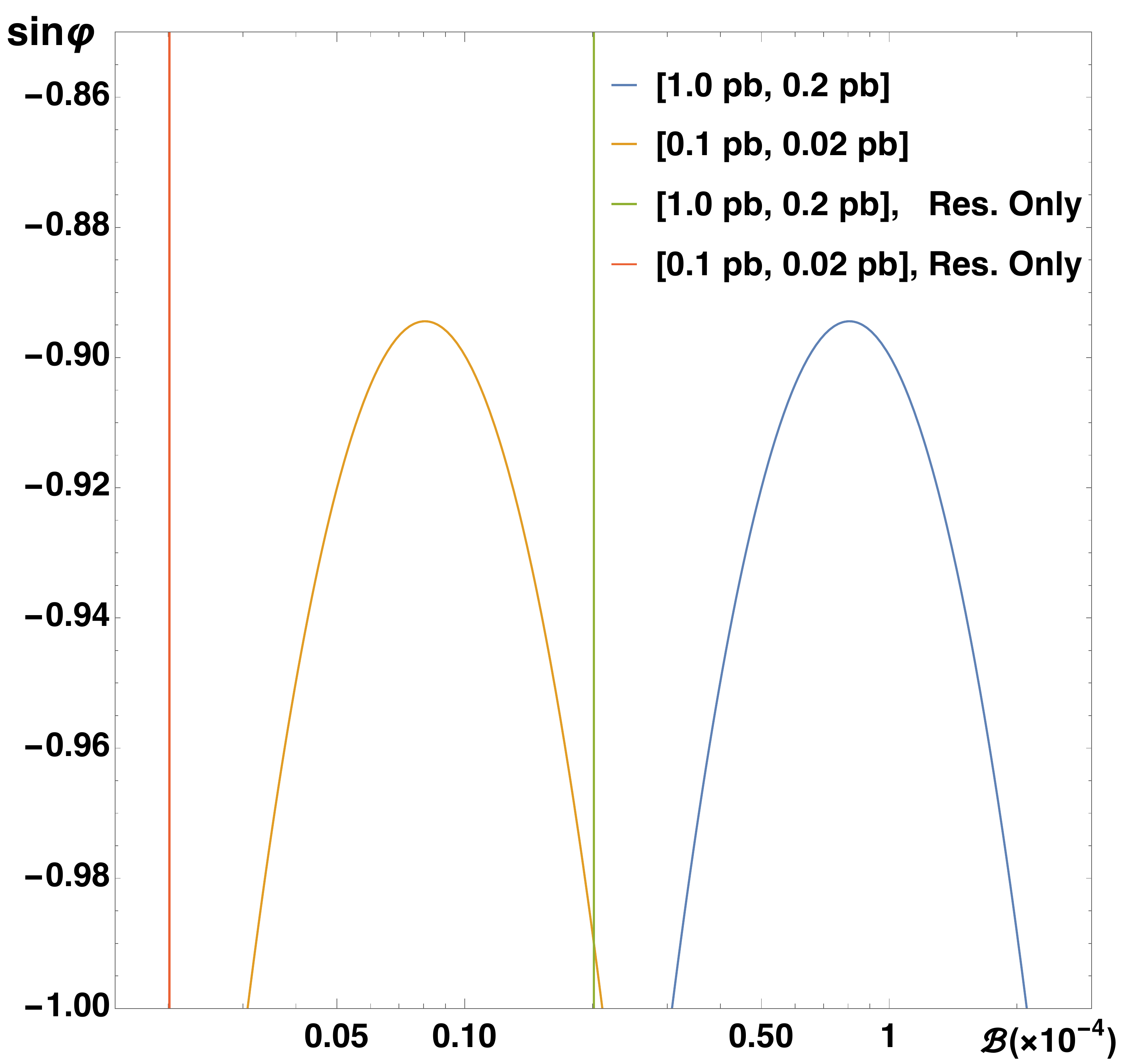}
    \caption{The two-dimensional functions of $\br^f$ and $\sin\varphi$ with 
    different $\sigma_{c}^f$ and $\sigma_{\rm tot}^f$ values estimated at the 
    $\psipp$ nominal mass. In the plots, the vertical lines represent
    the branching fraction calculated with interference term neglected (No Int.)
    or both the interference and continuum terms neglected (Res. Only).}
    \label{fig:br-vs-phi-psi3770}
\end{figure}

As mentioned earlier, the interference term will introduce a deviation, either positive 
or negative, to the continuum cross section at the resonance peak position.
In both cases, neglecting the interference term, as has been done in previous 
measurements~\cite{BESIII:2013ujm,BES:2007zan,CLEO:2005tkm,CLEO:2005zrs,Belle:2013hkg,Belle:2009rfa},
will lead to imprecise branching fractions, as discussed in Ref.~\cite{Wang:2005sk}. 
If $\sigma_{\rm tot}^f$ is larger than $\sigma_c^f$, the true branching fraction 
determined with an interference effect taken correctly into account
could be larger or smaller than the one determined by simply subtracting the continuum contribution, 
depending on the value of $\varphi$. If $\sigma_{\rm tot}^f$ 
is smaller than $\sigma_c^f$, neglecting the interference term only will result 
in an unphysical branching fraction value, while neglecting both will lead to a smaller 
branching fraction. In Fig.~\ref{fig:br-vs-phi-psi3770}, 
two-dimensional functions of $\br^f$ and $\sin\varphi$ with typical 
$[\sigma_{c}^f,~\sigma_{\rm tot}^f]$ values at $\psipp$ peak position
are shown, the cases where $\sigma_{\rm tot}^{f}$ are larger or smaller
than $\sigma_{c}^{f}$ are displayed separately. In both cases, the vertical lines in the plots
represent the branching fractions when the
interference term or both the interference and continuum terms are neglected.
Although the absolute difference between the branching fractions calculated with 
or without the interference and continuum contribution depends on the relative phase,
it is very significant in most of the parameter space. 

\section{Additional experimental effects}\label{ch:discussion}

In the above discussion, the radiative correction and energy spread of the colliding 
beams are not considered to make the discussions clear and simple. We prove here 
that these two effects do not affect the conclusions above. 

Taken these effects into account, the observed cross section 
can be written as
\begin{equation}\label{eq2:observe-cs}
    \sigma_{\rm exp}^f(s)=\int_{0}^{1-s_{m}/s}dx\int_{0}^{\infty}d\sqrt{s'}F(x,s')
    \cdot\sigma^f(s'(1-x))\cdot G(\sqrt{s},\sqrt{s'}).
\end{equation}
Here $x=1-s_{\rm eff}/s$ and $\sqrt{s_{\rm eff}}$ represents the effective c.m. 
energy after losing energy due to photon emission, $\sqrt{s_{m}}$ is the cutoff 
of $\sqrt{s_{\rm eff}}$ for the final state system and should be at least as large 
as the invariant mass of the final state system, $\sqrt{s}$ is the nominal c.m. 
energy, and $\sqrt{s'}$ is the actual c.m. energy which differs from the nominal 
one due to beam energy spread. The radiator function $F(x,s)$ is calculated 
with a precision of $0.1\%$ in Ref.~\cite{radiator-function}. 
$G(\sqrt{s},\sqrt{s'})=\frac{1}{\sqrt{2\pi}\Delta}e^{-\frac{(\sqrt{s}-\sqrt{s'})^2}
{2\Delta^2}}$ is the 
beam energy spread function where $\Delta$ stands for the c.m. energy spread. 

\begin{figure}[htbp]
    \centering
    \includegraphics[width=0.48\textwidth]{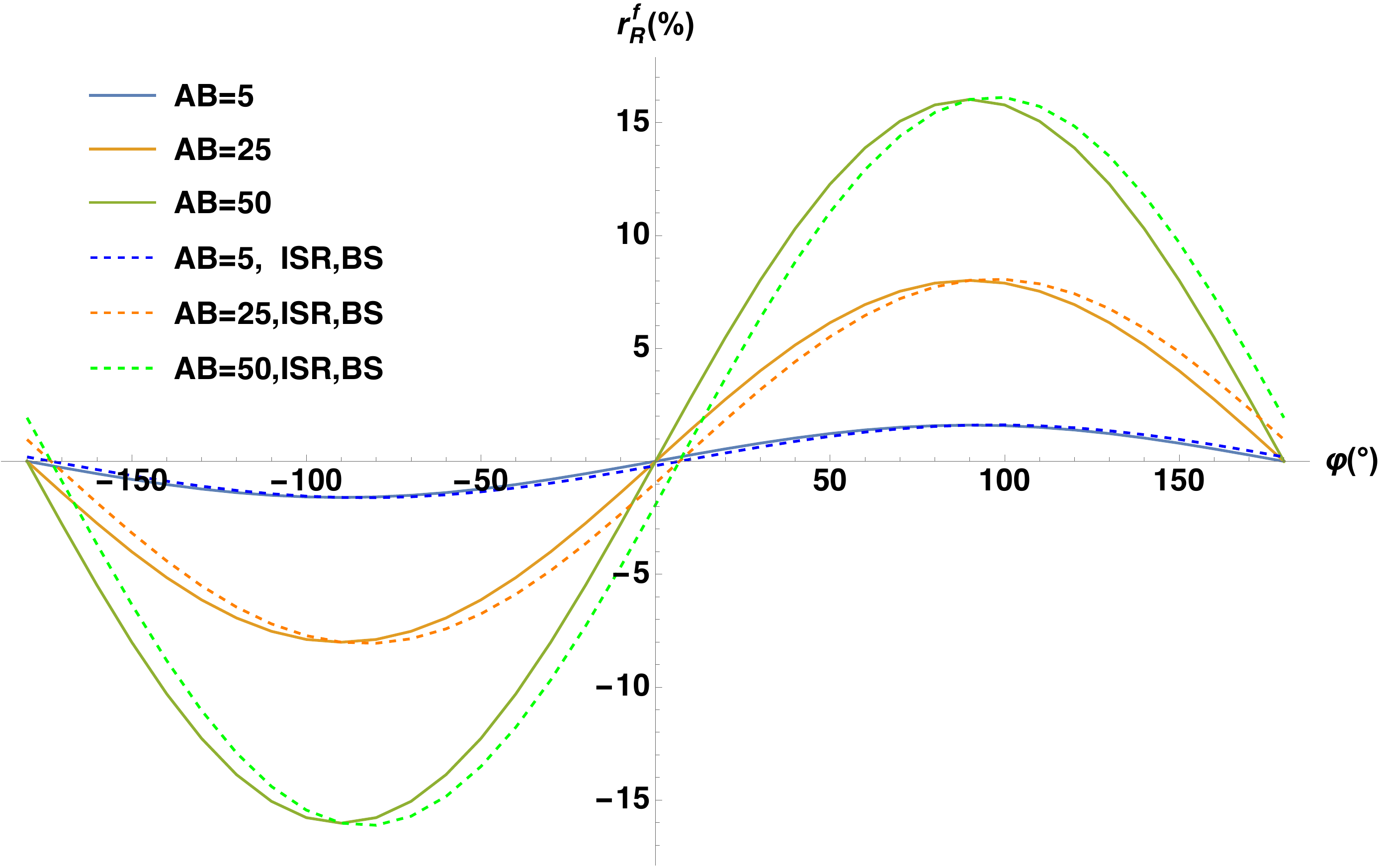}
    \includegraphics[width=0.48\textwidth]{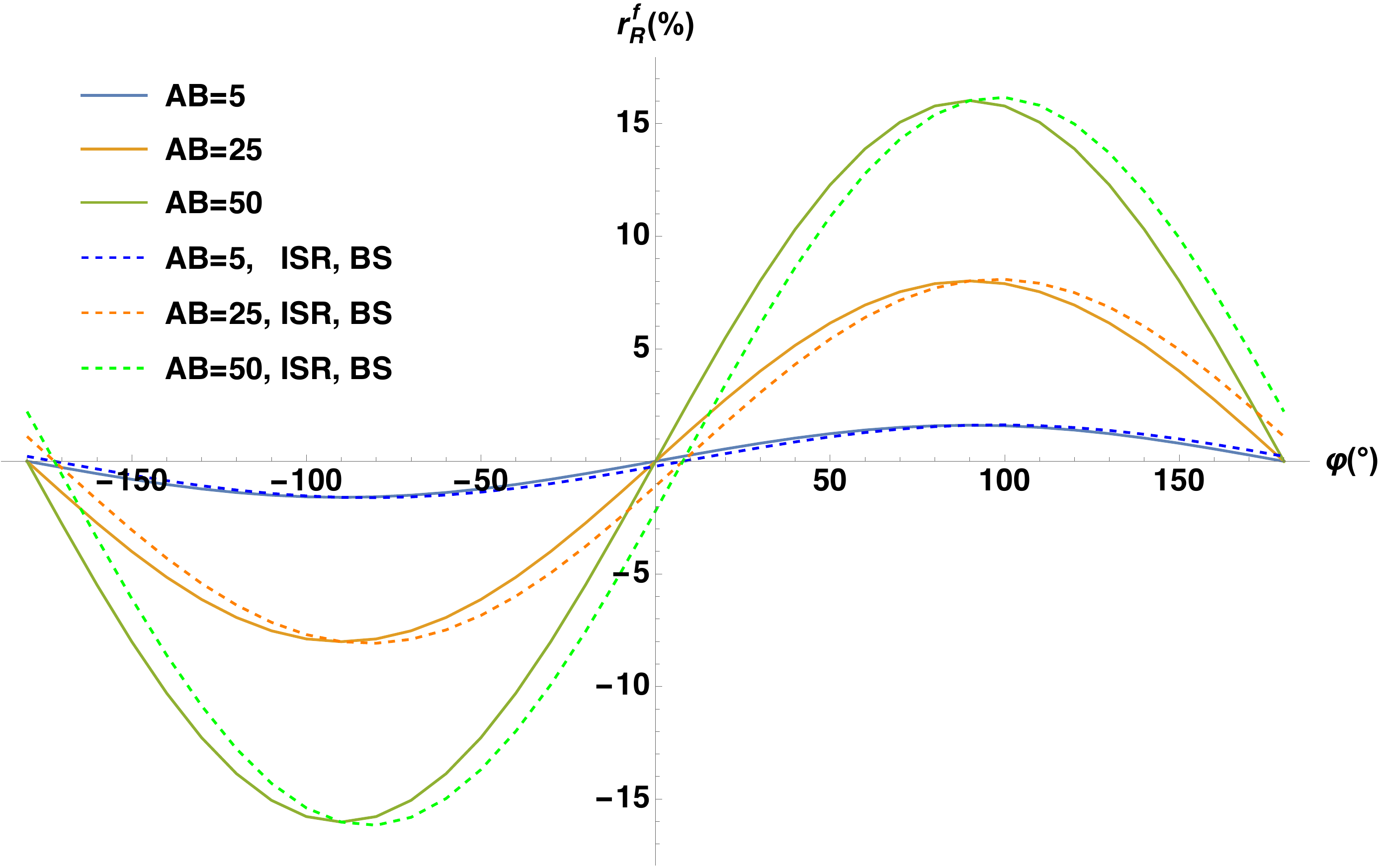}\\
    \includegraphics[width=0.48\textwidth]{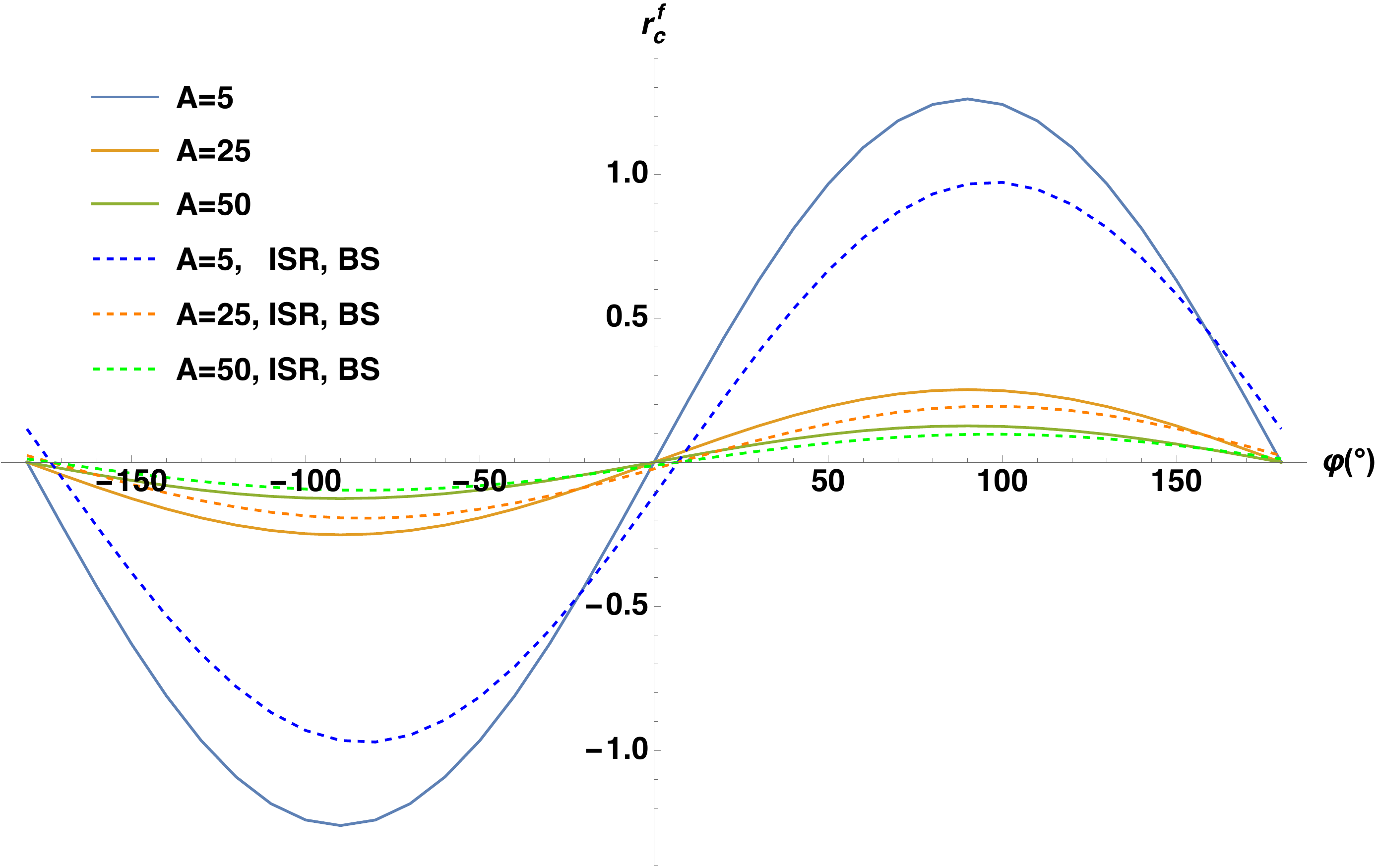}
    \includegraphics[width=0.48\textwidth]{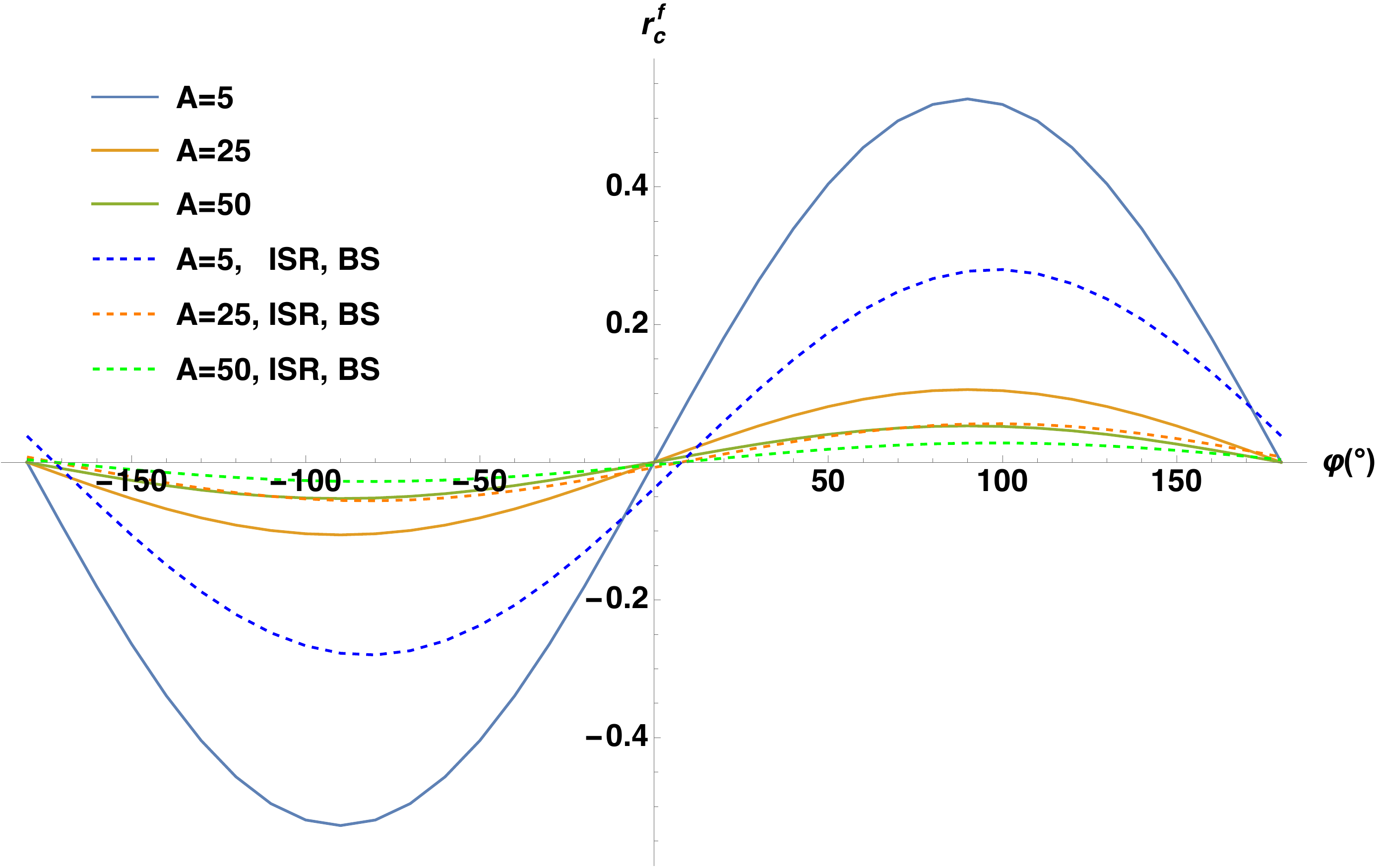}
\caption{The dependence of $r_{R}^{f}$ (top row) and $r_{c}^{f}$
(bottom row) on $\varphi$ 
with different $AB$ before (solid lines) and after (dashed lines) considering 
radiative correction 
(ISR) and beam energy spread (BS). The top left plot is at $\jpsi$ peak position, 
the top right is at $\upsi$ peak position, the bottom left is at $\psipp$ peak 
position, and the bottom right is at $\Upsilon(4S)$ peak position. }
    \label{fig:r-jpsi-ISR-BS}
\end{figure}

\begin{figure}[htbp]
    \centering
    \includegraphics[width=0.65\textwidth]{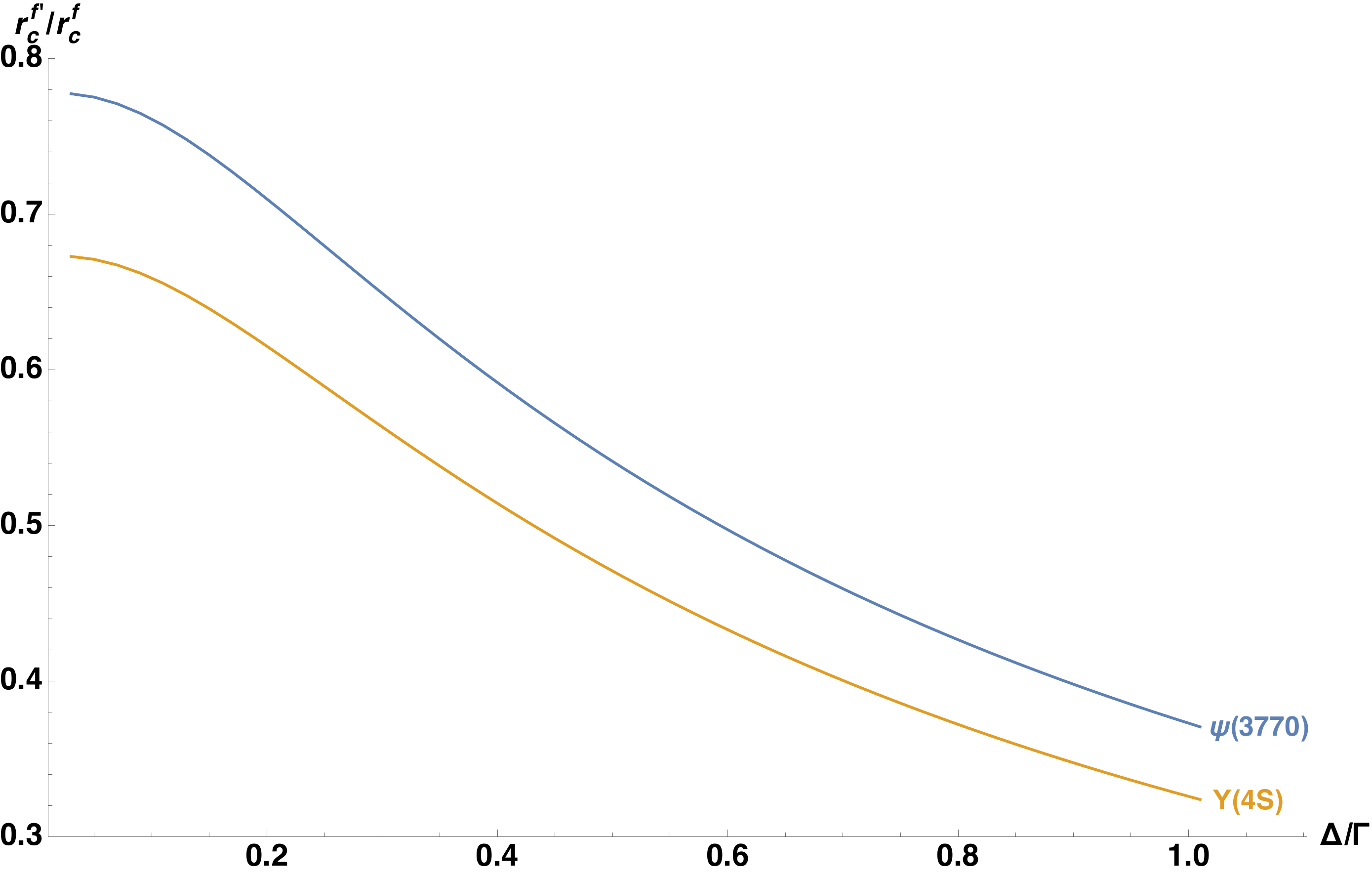}
\caption{The reduction of $r_{c}^{f}$ from the radiative correction and energy spread
for $\psipp$ and $\Upsilon(4S)$ as a function of $\Delta/\Gamma$, illustrated by $r_{c}^{f\prime}/r_{c}^{f}$
with $r_{c}^{f\prime}$ stands for that after considering the two effects.
}
 \label{fig:ratio-rc-as-BS}
\end{figure}

At the BESIII experiment, the $\Delta$ values for $\jpsi$, $\psip$, and $\psipp$ are 
$0.8$, $1.3$, and $1.4~\mev$, respectively. At the Belle II experiment, the typical 
$\Delta$ value is about $\sqrt{2}\cdot 5~\mev$. Using~\Cref{eq2:observe-cs} and 
replacing $\sigma^{f}$ with~\Cref{eq:born-cs-expand}, the dependence of 
$r_{R}^{f}$ and $r_{c}^{f}$ on $\varphi$ after considering 
radiative correction and energy spread can be obtained. The results shown in 
Figs.~\ref{fig:r-jpsi-ISR-BS} and~\ref{fig:ratio-rc-as-BS} are 
calculated with the integral over $dx$ done in the range of $[0,~0.2]$ and 
the continuum cross section $\sigma_{c}^{f}(s)$ assumed to be proportional to $1/s$.

The $r_{R}^{f}$ distributions with different $AB$ at $\jpsi$ and $\upsi$ masses and 
$r_c^{f}$ distributions with different $A$ ($B$ set to values from
Table~\ref{tab:born-cs-narrow-res})
at $\psipp$ and $\Upsilon(4S)$ masses are shown in Fig.~\ref{fig:r-jpsi-ISR-BS}. 
The distributions for other resonances are similar. 
In both $r_{R}^{f}$ and $r_{c}^{f}$ distributions, there is a shift along $\varphi$
that is caused by radiative correction. For a narrow resonance, the size of 
$r_{R}^{f}$ is almost the same before and after taking radiative correction and energy 
spread into account. For broad resonances, the two effects reduce
the maximum of $r_c^{f}$ by $23\%$ for $\psipp$ at the BESIII experiment
and $47\%$ for $\Upsilon(4S)$ at the Belle II experiment. The size of the reduction 
depends on the resonance parameters, and $\Delta$, does not depend on the value of $A$.  
Figure~\ref{fig:ratio-rc-as-BS} shows the ratio of $r_c^{f}$ calculated with or 
without radiative correction and energy spread, $r_{c}^{f\prime}/r_c^{f}$, as a function of $\Delta/\Gamma$.

\section{Summary}

In this paper, the importance of the interference between the continuum and 
resonance amplitudes in the measurement of branching fractions of the decays 
of vector quarkonia at $\EE$ colliders is discussed. 
The exact formula that can be used to estimate the size of the interference effect 
is investigated. Although the absolute contribution from the continuum process 
is negligible for narrow resonances, the interference contribution can be at a few percent level, 
depending on the final states. Whereas for the broad resonances, 
the interference effect could be even more significant and needs to be taken into account 
properly to avoid wrong interpretation of the data. 

With the currently available data samples at the BESIII experiment and the
expected data samples at the Belle II experiment, the precision 
of the branching fractions is expected to be at a few percent or better level, 
so mishandling the interference effect will lead to systematic 
bias with a much larger size compared to the statistical uncertainty and comparable 
to or even larger than all the other sources of the systematic uncertainties. 
The optimal solution to this problem is to change the data taking strategy:
accumulate data at no less than three different energies in the vicinity of a 
resonance, and measure the continuum and the resonance decay amplitudes
together with the relative phase between them.

At colliders planned for the future, such as the super-tau-charm factories 
STCF in China~\cite{STCF} and SCT in Russia~\cite{SCT}, and the 
super-$\jpsi$ factory~\cite{super-jpsi}, the design luminosity is expected to be 
hundreds of times higher than the current tau-charm factory. 
It will be even more crucial to handle the interference properly as 
the precision of the measurements can be further improved with a few orders of
magnitude larger data samples. 

\acknowledgments

This work is supported in part by National Key Research and Development 
Program of China under Contract No. 2020YFA0406300, 
Joint Large-Scale Scientific Facility Funds 
of the NSFC and CAS under Contract No. U203210, 
and NSFC under contracts No. 11961141012 and No. 11835012.

%

\end{document}